# Integrated electro-optically tunable narrow-linewidth III-V laser

Yiran Zhu,[1,2] Shupeng Yu,[3,4] Zhiwei Fang,[2,*] Difeng Yin,[2] Jian Liu,[2] Zhe Wang,[2] Yuan Zhou,[3,4] Yu Ma,[3,4] Haisu Zhang,[2] Min Wang,[2] And Ya Cheng[1,2,3,5,6,7,8,*]

*Corresponding Author: E-mail: zwfang@phy.ecnu.edu.cn; ya.cheng@siom.ac.cn

[1]State Key Laboratory of Precision Spectroscopy, East China Normal University, Shanghai 200062, China.
[2]The Extreme Optoelectromechanics Laboratory (XXL), School of Physics and Electronic Science, East China Normal University, Shanghai 200241, China.
[3]State Key Laboratory of High Field Laser Physics and CAS Center for Excellence in Ultra-intense Laser Science, Shanghai Institute of Optics and Fine Mechanics (SIOM), Chinese Academy of Sciences (CAS), Shanghai 201800, China.
[4]Center of Materials Science and Optoelectronics Engineering, University of Chinese Academy of Sciences, Beijing 100049, China.
[5]Collaborative Innovation Center of Extreme Optics, Shanxi University, Taiyuan, Shanxi 030006, People's Republic of China.
[6]Collaborative Innovation Center of Light Manipulations and Applications, Shandong Normal University, Jinan 250358, People's Republic of China.
[7]Hefei National Laboratory, Hefei 230088, China.
[8]Joint Research Center of Light Manipulation Science and Photonic Integrated Chip of East China Normal University and Shandong Normal University, East China Normal University, Shanghai 200241, China.

**Abstract**: We demonstrate an integrated electro-optically tunable narrow-linewidth III-V laser with an output power of 738.8 μW and an intrinsic linewidth of 45.55 kHz at the C band. The laser cavity is constructed using a fiber Bragg grating (FBG) and a tunable Sagnac loop reflector (TSLR) fabricated on thin film lithium niobate (TFLN). The combination of the FBG and the electro-optically tunable TSLR offers the advantages of single spatial mode, single-frequency, narrow-linewidth, and wide wavelength tunability for the electrically pumped hybrid integrated laser, which features a frequency tuning range of 20 GHz and a tuning efficiency of 0.8 GHz/V.

## 1. Introduction

The tunable narrow-linewidth lasers have been widely used in high-speed optical communication, laser cooling, and precision metrology thanks to the wide frequency tunability, high coherence, and low phase noise. Currently, most tunable narrow-linewidth lasers are constructed based on solid-state lasers, fiber lasers, and semiconductor lasers, as these lasers are of high power efficiency and easy maintenance [1,2]. Different approaches have been used to realize tunable narrow-linewidth lasers, although in principle, these approaches all employ highly spectroscopically resolvable components combined with the laser cavities to achieve frequency filtering and linewidth suppression. For instance, in solid-state lasers non-planar ring oscillators are frequently utilized, while in the fiber lasers the choice changes to the distributed Bragg grating (FBG) inscribed within the fiber [3,4]. Alternatively, in the semiconductor lasers, a large external cavity structure incorporating wavelength-sensitive components such as diffraction gratings and lenses are typically involved [5,6]. Nevertheless, in these bulk systems, issues related to large footprint, high cost of packaging, and low tuning rate of laser wavelength are inevitably.

In recent years, the integration of external cavity constructed with photonic integrated circuits (PICs) and semiconductor gain chips has attracted tremendous attention for achieving compact, low-power consumption, and highly stable narrow-linewidth semiconductor lasers [7-16]. The miniaturized lasers use Vernier microring resonators, distributed Bragg reflectors, Sagnac loop mirrors, and Mach-Zehnder Interferometers (MZIs) to form the laser cavity, providing the necessary optical feedback for suppressing the linewidth and expanding the wavelength tuning range. Such hybrid integrated tunable semiconductor lasers are of small footprint and high stability, in which the external cavities have been built upon various material platforms including silicon dioxide ($SiO_2$), silicon (Si) and silicon nitride ($Si_3N_4$) using advanced microfabrication technologies [10-12]. In addition, external cavities based on TFLN have recently been intensively investigated due to its characteristic nonlinear optical effect, ultra-wide optical and electrical bandwidth, and ultra-high electro-optic modulation rate and efficiency [17–20]. Such advantages have led to the recent demonstrations of integrated tunable, frequency conversion continuous-wave TFLN laser as well as mode-locked lasers [21–25]. The main efforts focus on promoting the laser output power, suppressing the linewidth, and expanding the wavelength tuning range. The ultimate goal is to achieve high performance tunable lasers with all the key parameters comparable or even better than that of the bulk laser systems.

Here, we demonstrate a hybrid integrated electro-optic tunable narrow-linewidth III-V laser by butt coupling a semiconductor optical amplifier (SOA) chip with a tunable Sagnac loop reflector (TSLR) and a fiber Bragg grating (FBG). On the output end of the integrated laser, the TSLR consisting of a Mach-Zehnder interferometer (MZI) and Sagnac loop mirror, which is fabricated on the TFLN substrate, is used to tune the laser wavelength as well as the reflectivity of the output window. The FBG fused to the output fiber of the SOA serves as a narrow-bandwidth reflector on the other end of the laser. We examine the performance of the hybrid integrated laser in terms of laser output power, linewidth, wavelength tuning range and tuning efficiency, as well as the light-current-voltage (L-I-V) curve.

## 2. Device Structure and Fabrication

**Figure 1**(a) illustrates the schematic structure of the hybrid TFLN/III-V laser, which is composed of an FBG, an SOA chip, and a TFLN external cavity chip. The Fabry-Perot (FP) laser cavity is defined between the high reflectivity FBG and the partially reflective TSLR on TFLN substrate. In the FP laser cavity, an InP-based SOA chip (Thorlabs) provide the electrically pumped optical gain. The laser wavelength can be electrically tuned using the TSLR on TFLN. The SOA chip is a C-band single angled facet (SAF) gain chip which uses a geometric design technique to further reduce the reflection at the front facet of the chip by curving the ridge waveguide at 5 degrees, by which eliminating back reflections that can create unwanted feedback into the laser cavity. In addition, reflection at the front facet is further reduced using anti-reflection coating. It is noteworthy that any residual reflection from the chip facet can have negative influence on the stability, output power, and spectral quality of the laser.

The FBG is connected at the rear facet of the SOA chip to achieve a high reflective end mirror of FP cavity. As shown in **Figure 1**(b), the output aperture width of SOA is 3 µm, thus the light from the SOA is butt-coupled into the TFLN chip through a tapered TFLN waveguide. Note that the input waveguide fabricated on the TFLN is also tilted by 5 degrees for achieving high coupling efficiency. The TFLN chip is fabricated on an 800-nm-thick TFLN substrates (800 nm TFLN/4.7 µm $SiO_2$/500 µm Si) using the photolithography assisted chemo-mechanical etching (PLACE) technique. More technical details of PLACE can be found in our previous work [26-28]. As shown in **Figure 1**(c), the TSLR consists of a MZI and a Sagnac loop mirror. The top width of waveguide is ~1.5 µm, and the etching depth of the ridge is 350 nm. The Sagnac loop mirror is of a radius of curvature of 500 μm, and the coupling length of the directional coupler (DC) is 330 µm and the gap width in the DC is ~3.3 µm. Electro-optical tuning of the refelctivity of TSLR is realized by connecting the Sagnac loop mirror with MZI. The MZI consists of two 2×2 3-dB DC and two phase-shift arms with push-pull electrodes, and the length of straigt arm waveguide arm is 5 mm. Meanwhile, gold (Au) electrodes are fabricated on the two sides of the straight waveguide arms according to the ground-signal-ground (GSG) configuration, giving rise to the necessary phase difference between the two arms of MZI. The width and length of the electrodes is 500 μm and 5 mm, respectively. The output port on theTFLN chip is coupled to a lensed fiber from which the laser signal is detected and analyzed, whilst the remaining light in the Sagnac loop mirror is reflected back into the SOA to provide the feedback for the stimulated amplification.

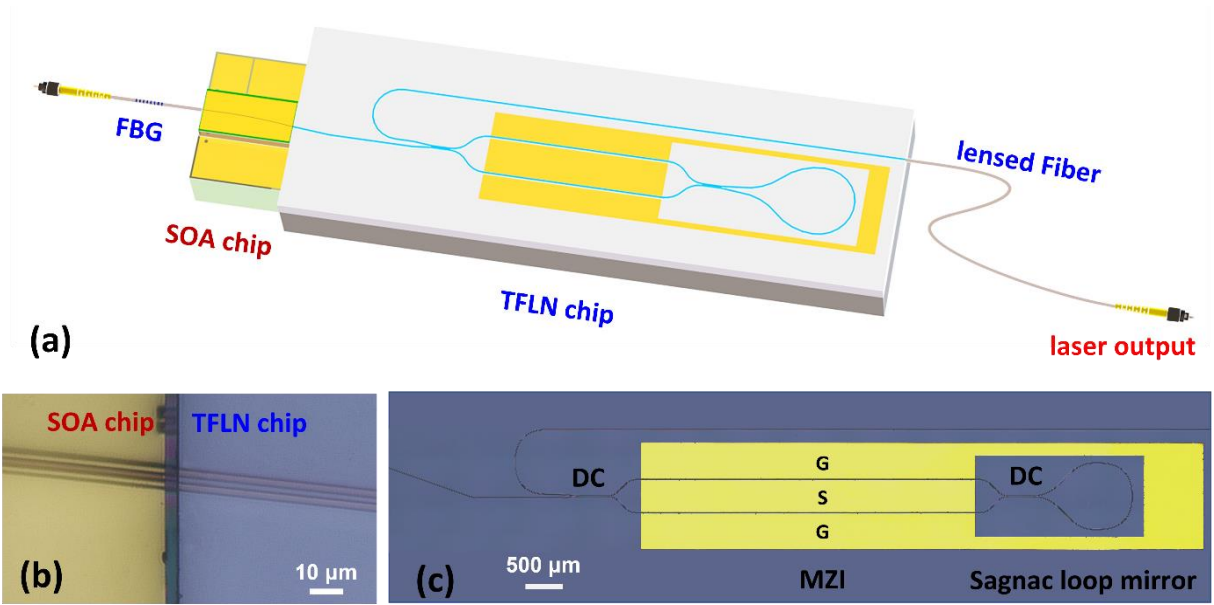

**Figure 1.** (a) Schematic view of the hybrid TFLN/III-V laser. (b) Optical microscope image of joint region between the TFLN chip and SOA chip. (c) Optical microscope image of the fabricated TSLR on TFLN chip.

## 3. Results and Discussions

As shown in **Figure 2** (a), we investigated the reflectivity properties of electro-optic TSLR by measuring the its optical transmission. In this experimental measurement setup, the tunable laser signal is generated from a C-band continuously tunable laser (CTL 1550, TOPTICA Photonics) and injected into the hybrid TFLN/III-V laser from the input end by butt coupling. At the beginning, the SOA does not generate optical gain when the power is off. Under this condition, we examine the output power from the TSLR as a function of the direct current voltage (IPMP250-1 L, INTERLOCK) applied between the ground and signal electrodes using two probes (ST-20-0.5, GGB). The output mode from the TSLR is imaged using a microscope with an infrared charge-coupled device (IR CCD) camera (InGaAs Camera C12741-03, Hamamatsu Photonics). **Figure 2**(b) shows the measured reflectivity of the fabricated TSLR as a function of the applied DC voltage, showing that the reflectivity can be tuned from near 0 to 84% when the driven voltage of the MZI phase shifter is varied from -80 V to 80 V with a tuning step of 5V. Ideally, the refelctivity can be tuned from 0 to 100% when the phase difference between the two arms of MZI changes from 0 to π/2. In reality, the reflectivity of TSLR cannot be tuned to absolute 100% due to the unavoidable fabrication error, which limits the ultimate extinction ratio of the MZI. The insets of **Figure 2**(b) show infrared images of the output mode of the hybrid TSLR at different voltages of -65V, 0V, and 45V. We also investigated the reflectivity properties of the FBG which is used in the hybrid TFLN/III-V laser. **Figure 2**(c) presents the reflection spectrum of the FBG, which shows a reflectivity of 99% at 1544.1 nm wavelength and a 3-dB bandwidth of 0.2 nm.

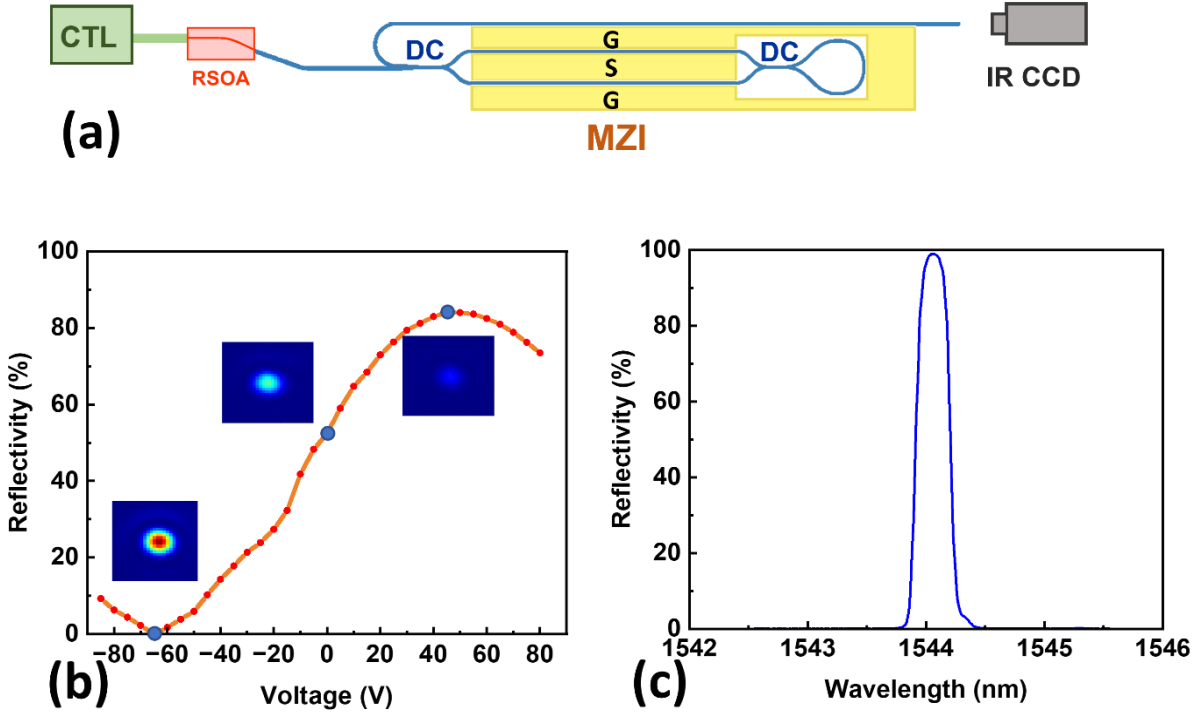

**Figure 2**. Experimental setup for measuring the reflectivity of TSLR under different DC voltages (CTL: continuously tunable laser; SOA: reflective semiconductor optical amplifier ; DC: directional coupler; MZI: Mach-Zehnder interferometer; G: ground; S: signal; IR CCD: infrared charge-coupled device). (b) Measurement results of reflectivity of TSLR with the varying DC voltage. The insets show the infrared images of the output mode from the TSLR at -65V, 0V, and 45V.(c) The reflection spectrum of FBG.

The experimental setup for characterizing the hybrid integrated laser is illustrated in **Figure 3**(a). In the experimental measurement setup, an optical spectrum analyzer (OSA: AQ6375B, Yokogawa) with a wavelength resolution of 20 pm is connected to the output port of the TSLR to collect and analyze the laser signal. A pair of pin probes are used to apply DC voltages on the gold contacts fabricated near the MZI waveguide arms for generating a phase shift in the MZI. **Figure 3**(b) shows is the experimental setup photographed by a digital camera, in which each components in the whole device can be clearly seen. **Figure 3**(c) shows the L-I-V curve of the laser when operating at the wavelength of 1544 nm, indicating a lasing threshold at an electric current of 130 mA.

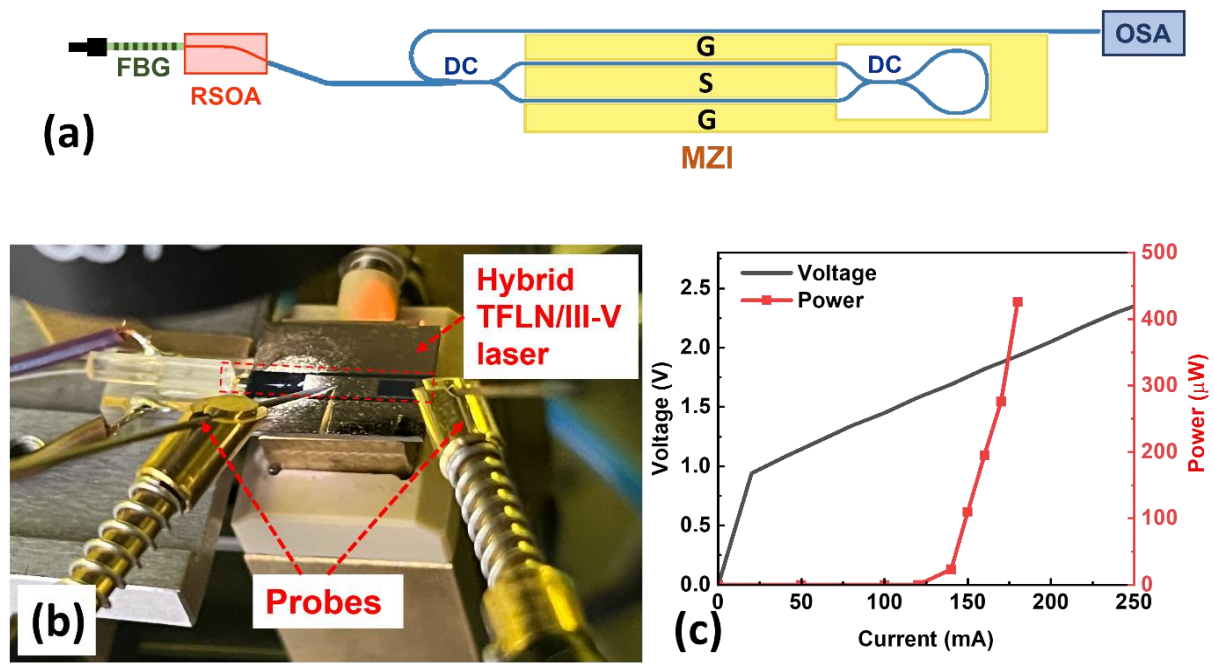

**Figure 3.** (a) Experimental setup for characterizing the hybrid TFLN/III-V laser (FBG: Fiber Bragg grating; SOA: reflective semiconductor optical amplifier ; DC: directional coupler; MZI: Mach-Zehnder interferometer; G: ground; S: signal; OSA: optical spectrum analyzer). (b) Photograph of the setup with a SOA butt-coupled to the TFLN chip. (c) The L-I-V curve of hybrid TFLN/III-V laser, red and black curve are L-I and I-V curves, respectively.

As shown in **Figure 4**(a), the laser performance in the wavelength range from 1540 to 1550 nm is characterized using the OSA. A single longitudinal mode peaked at 1544 nm wavelength is generated due to the narrow bandwidth (0.2 nm) of FBG. When the applied voltage is 45 V at which the reflectivity of TSLR is highest, a maximum output power of 738.8 µW at 1544 nm wavelength is measured. As shown in the inset of **Figure 4**(a). the generated laser is also of the fundamental spatial mode form the image captured by the IR CCD camera. The linewidth of the hybrid TFLN/III-V laser is measured using a home-built delayed self-heterodyne interferometer. In the measurement setup, a single-mode fiber with a length of 5 km was used in one optical path, and another optical path is incorporated with an acousto-optic modulator which shifted the light frequency by 200 MHz. The extracted line shape can be fitted using a Lorentz function centered at the frequency of 200 MHz as shown in **Figure 4**(b), indicating an intrinsic linewidth of 45.55 kHz.

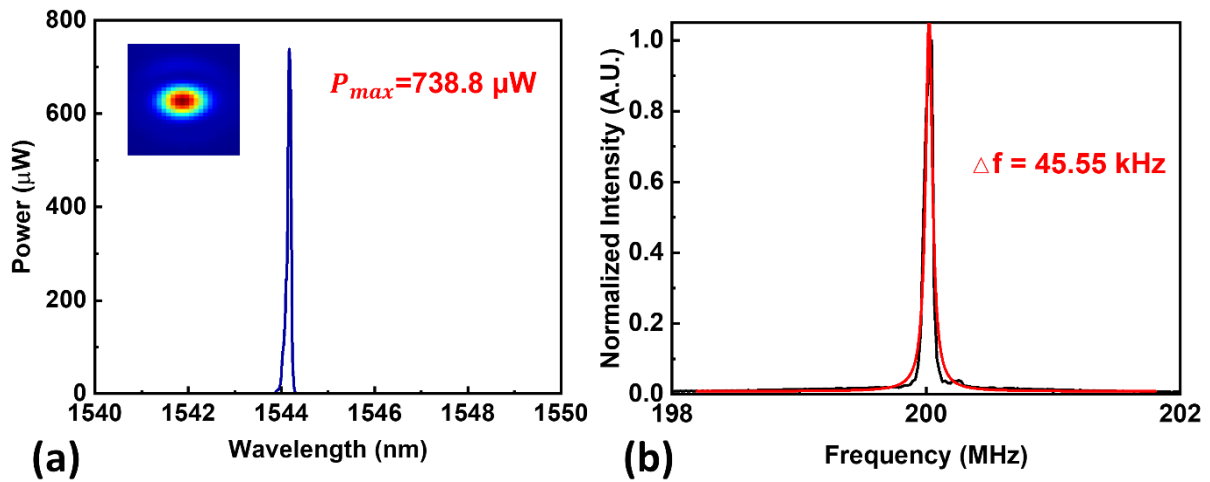

**Figure 4** (a) The spectrum of the on-chip maximum output single mode laser power of 738.8 µW, the resolution the OSA is 20 pm, the inset shows an infrared image of the output port of the hybrid TFLN/III-V laser. (b) The linewidth of 45.55 kHz is extracted by a delayed self-heterodyne interferometric measurement.

The laser wavelength can be tuned by varying the applied direct current voltage. **Figure 5**(a) shows the superimposed lasing spectra captured using the OSA under different direct current voltages separated with an interval of 25V. The laser wavelength can be tuned from 1544.01 nm to 1544.17 nm, which corresponds to 20 GHz tuning range. The tuning range is within the bandwidth of the FBG (3-dB bandwidth of 0.2 nm at central wavelength 1544.1 nm). **Figure 5**(b) shows the linear dependence of the lasing frequency on the applied direct current voltage, from which a tuning efficiency of 0.8 GHz/V can be derived.

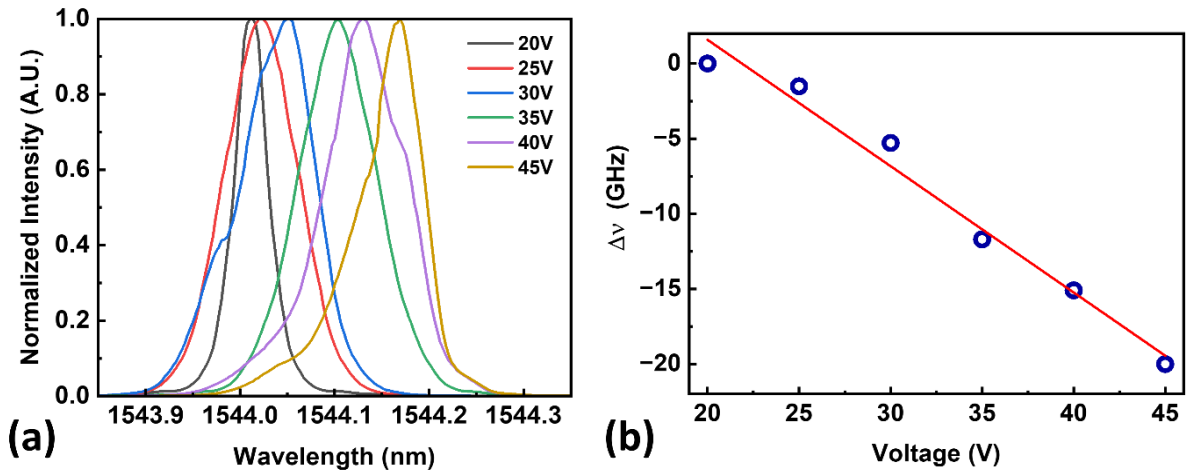

**Figure 5.** (a) Measured laser spectra under different direct current voltages applied to the MZI. The spectral resolution is 20 pm as limited by OSA. (b) Linear laser-frequency detuning as a function of signal voltage, showing a tuning efficiency of 0.8 GHz/V.

## 4. Conclusion

In conclusion, we demonstrate an on-chip hybrid integrated TFLN/III-V laser of an output power of 738.8 μW and intrinsic linewidth of 45.55 kHz in the C band. This laser also exhibits a frequency tuning range of 20 GHz and a tuning efficiency of 0.8 GHz/V. The laser is unique in its hybrid external cavity design which combines a TFLN TSLR and an FBG, both of which can be easily integrated to the SOA chip. The combination of the two makes the filtering of single frequency out of the laser cavity straightforward and allows for high-precision wavelength tuning thanks to the high reflective Sagnac loop reflector. It seems that the tuning range of this approach would suffer from the limited bandwidth of the FBGs, while this difficulty can be easily overcome by fusing an array of FBGs connected to an on-chip multi-channel optical switch to cover a broad wavelength range. Future efforts will focus on the optimization of the coupling efficiency of TFLN external cavity with the SOA chip to boost the output power and improving the Q-factor of Sagnac loop reflector to narrow down the linewidth. In addition, high-speed wavelength modulation of the laser will be investigated by comparing various design schemes such as electro-optic tuning of the Sagnac loop mirror or incorporating a high-speed phase modulator based on TFLN waveguide into the laser cavity. Last but not the least, by replacing the FBG reflector with a high-reflectivity coating on the input end of the SOA, this hybrid TFLN/III-V laser structure can be used for creating on-chip Q-switched lasers and mode-locked lasers.

## Acknowledgements

National Key R&D Program of China (2019YFA0705000, 2022YFA1404600, 2022YFA1205100), National Natural Science Foundation of China (12274133, 12004116, 12104159, 12192251, 11933005, 12134001, 61991444), Science and Technology Commission of Shanghai Municipality (21DZ1101500), Shanghai Sailing Program (21YF1410400). Innovation Program for Quantum Science and Technology (2021ZD0301403), the Fundamental Research Funds for the Central Universities (East China Normal University). We thank SJTU-Pinghu Institute of Intelligent Optoelectronics for chip packaging.